\documentclass[twocolumn,showpacs,prb,aps,epsf,amsmath,amssymb]{revtex4}
\usepackage{graphicx}

\begin{document}

\title{
Separating parallel conduction from two-dimensional magnetotransport in high mobility InP/InGaAs MOCVD-grown heterostructures
}

\author{Yang Tang, Chuanle Zhou, M. Grayson}
\affiliation{
Electrical Engineering and Computer Science, Northwestern University, Evanston, IL 60208 USA}

\begin{abstract}
In this Letter, four-point magnetotransport of high mobility InGaAs/InP heterointerfaces is measured from 1.6 K to 300 K and from 0 to 15 T, and an analysis is shown whereby the mobility and density of the two-dimensional (2D) accumulation layer can be separately characterized from that of the parallel conducting dopant layer over all but a small intermediate temperature range.
Standard magnetotransport regimes are defined as the temperature increases from 1.6 K to 300 K, namely quantum Hall (QH), Shubnikov de Haas (SdH), and Drude regimes (D), and in the QH and D regimes different analyses are applied to deduce densities and mobilities of both layers separately. Quantitative conditions for the intermediate SdH regime are defined, within which both QH and D analyses fail.  The density and activation energy of unintentional donors at the InP epilayer/substrate interface is deduced.
At base temperature, QH minima are resolved down to $B = 0.4$\,T at $\nu = 20$, revealing a mobility of $\mu = 160,000 \, \mathrm{cm}^2/\mathrm{Vs}$. 
The 2D system maintains this high mobility up to at least 40 K in this high quality structure.
\end{abstract}


\maketitle

High efficiency quantum cascade lasers require high quality MOCVD InP/InGaAs heterostructure growth, which can be characterized by examining electrical transport at a single InP/InGaAs interface. Such characterizations of high mobility two-dimensional electron gas (2DEG) heterostructures are often obscured by parallel conduction arising from intentional or unintentional doping from an undepleted bulk \cite{Hiyamizu,Schu,Hurd,Sven,Kane}, a second occupied subband \cite{Grayson}, an overdoped modulation doping layer \cite{Grayson}, or unintentional interface defects and impurities which can act as dopants in a localized layer. 
Various parallel conduction models have been used to characterize parallel conducting channels in both quantized \cite{Grayson} and non-quantized \cite{Kane,Hurd} magnetotransport of high mobility systems. But to date, experimental characterizations of InP/InGaAs heterostructures are lacking, typically because nominally undoped bulk layers nonetheless have moderately high background doping levels, and because interface defects and impurities induce localized donor states at the epilayer interface of unknown density. To date, parallel conduction characterizations have been performed principally at a single temperature, and temperature dependent characterizations for parallel channels over a wide temperature range was rarely reported.

In this paper, we present an analysis of magnetotransport data for an InP/InGaAs heterostructure from 1.6 K to 300 to separately extract the carrier density and mobility of the parallel conducting Drude layer from the high mobility layer. Three magnetotransport regimes are labelled according to the conductivity regime of the high mobility layer, namely quantum Hall (QH), Shubnikov de Haas (SdH) and Drude (D). 
In the high temperature D range, a new technique is presented whereby the magnetic field dependence of the conductivity over a large field range permits separation of the two parallel conduction parameters.   In low temperature QH range, the conductivity in the resistance minima reveals the conduction parameters of the parallel layer. The high quality of the heterostructure growth allows us to identify epilayer interface states as the principle source of the parallel channel in this InP/InGaAs system, and a quantitative model for identifying the heterolayer conduction band offset from the data is discussed.

%
\begin{figure}
	\includegraphics[width=\columnwidth]{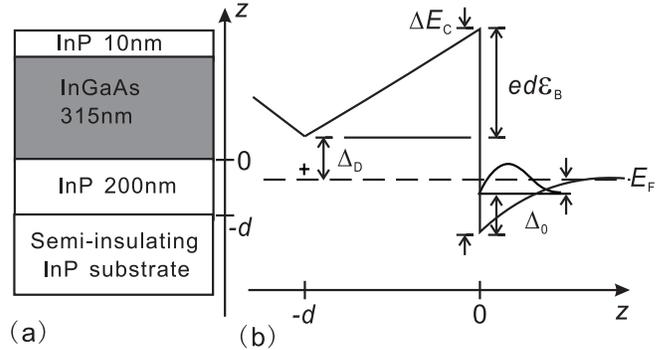}
    \caption{(a) Schematic configuration of the InP/InGaAs heterojunction.
    (b) Conduction band structure at the InP/InGaAs interface. $E_F$, $\Delta _D$, and $\Delta_0$ are the fermi level, binding energy of interfacial dopants, and the first electron subband in the quantum well respectively. The cross stands for ionized donors and the wavefunction is corresponding to the first subband.}
	\label{fig:structure}
\end{figure}
The sample is grown by metalorganic chemical vapor deposition \cite{razeghi,razeghi2} and the structure is in Fig.~\ref{fig:structure}(a). The composition of InGaAs layer is In$_{0.532}$Ga$_{0.467}$As. Neither the InP nor  InGaAs layer was intentionally doped. Because of the high growth quality, it is known that the background doping of the bulk layers is at most $10^{15} \, \mathrm{cm}^{-3}$. The interface between InP and semi-insulating InP substrate is known to be a possible additional source of donors for creating a parallel conducting channel, which is expected to dominate parallel conduction in such high growth quality structures. Panel (b) of Fig.~\ref{fig:structure} shows the conduction band diagram under the assumption that the latter is the dominant source of dopants.

Transport properties of the sample are measured using ac four point method in van der Pauw geometry to eliminate the influence of the contacts resistance\cite{vanderPauw}, and the film is patterned in a clover-shape. Standard lock-in techniques are used for the current source and voltage measurements. The sample is tested in an Oxford VTI helium gas flow cryostat, varying the temperature from 300 K down to 1.5 K. A superconducting magnet supplies a magnetic field up to 15 T.

As previously described, the temperature range is divided into three regimes according to the behavior of the high mobility layer in the highest magnetic fields, namely quantum Hall (QH), Shubnikov de Haas (SdH) and Drude (D) regimes.  The conductance parameters of the high mobility quantum confined layer will be labelled Q, and those of the parallel layer with the symbol $||$.

\begin{figure}
	\includegraphics[width=\columnwidth]{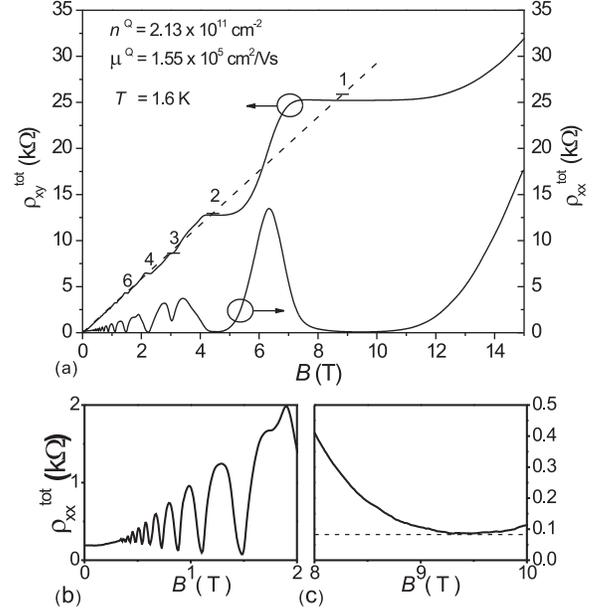}
    \caption{(a) Plot of measured $\rho_{xy}^{tot}$ and $\rho_{xx}^{tot}$ at 1.6 K. Quantized Hall conductance values $h/\nu e^2$ are indicated with short lines and indexed by the filling factor $\nu$. Carriers are $n$-type as revealed by the sign of Hall voltage. (b) Shubnikov de Haas oscillations in low magnetic field. (c) $\rho_{xx}^{tot}= 84 \, \Omega$ at the minimum  $\nu=1$.     }
	\label{fig:QHE}
\end{figure}

We begin sample characterization at base temperature in the QH regime and follow the notation of Grayson and Fischer \cite{Grayson} for modeling parallel conduction in this regime. At 1.6 K, magnetic field $B$ dependent measurement of the total longitudinal sheet resistance $\rho_{xx}^{tot}$ and transverse Hall resistance $\rho_{xy}^{tot}$ shows oscillations in $\rho_{xx}^{tot}$ which do not quite reach zero in Fig.~\ref{fig:QHE}(c), and plateaus in $\rho_{xy}^{tot}$ which likewise do not quite reach the nominal quantized plateau values in Fig.~\ref{fig:QHE}(a).  Both behaviors are indications of possible parallel conduction.
In the low $B$ region a single series of SdH oscillations periodic in 1/$B$ begins around 0.5 T with no evidence of beating in Fig.~\ref{fig:QHE}(b), revealing that only one type of high-mobility carrier exists in this system, and eliminating the possibility that a quantized second subband in the quantum well is the source of any parallel conduction \cite{razeghi3}.
Thus we can safely assume that 1.6 K is in the parallel conducting QH regime and follow the analysis of Ref.\,\cite{Grayson}. The carrier concentration of the quantum confined 2DEG $n^Q$ can be calculated from quantum oscillations to be $n^{Q} = 2.1 \times 10^{11} \, \mathrm{cm}^{-2}$, and the mobility can be calculated assuming the conductivity of the parallel channel is negligible compared to that of the 2DEG.  The result reveals a high mobility of $160,000 \, \mathrm{cm}^2/\mathrm{Vs}$. According to Ref.\,\cite{Grayson}, 
the small carrier density and mobility in the parallel channel $n^{||}$ is
\begin{equation}
\label{equ:n||}
{
n^{||}\approx n^Q \left( \frac{h/\nu e^2}{\rho_{xy}^{tot}}-1 \right)
}
\end{equation}
\begin{equation}
\label{equ:mu||}
{
\mu^{||}=\frac{\sigma_{xx}^{tot}}{n^{||}e}=\frac{1}{n^{||}e}\frac{\rho_{xx}^{tot}}{(\rho_{xx}^{tot})^2+(\rho_{xy}^{tot})^2},
}
\end{equation}
where $\rho_{xx}^{tot}$ and $\rho_{xy}^{tot}$ are the measured longitudinal and transverse sheet resistivity, $\nu$ the filling factor, $e$ the electron charge and $h$ the Planck constant. These equations result in $n^{||}=5.9 \times 10^9 \, \mathrm{cm}^{-2}$ and $\mu^{||}=140 \, \mathrm{cm}^2/\mathrm{Vs}$. The two orders of magnitude lower density confirms the assumption that parallel channel conduction is comparatively small.

Equations \ref{equ:n||} and \ref{equ:mu||} are only valid when the high mobility layer is in the QH regime, with longitudinal and transverse conductivities $\sigma_{xx}^{Q}=0$ and $\sigma_{xy}^{Q}=ne^2/h$, respectively.  At higher temperatures $\sigma_{xx}^{Q} \neq 0$, and the high mobility layer enters the SdH regime wherein no analysis can separate the the conductances of the two layers.
When $\sigma_{xx}^{Q}$ is finite, the 
parallel channel transverse resistance $\rho_{xy}^{||}$  becomes
\begin{equation}
\label{equ:RxySdH}
{
\rho_{xy}^{||}=\frac{|\boldsymbol\rho^{tot}|(\rho_{xy}^{tot}-|\boldsymbol\rho^{tot}|\nu e^2/h)}{(\rho_{xx}^{tot}-\sigma_{xx}^{Q}|\boldsymbol\rho^{tot}|)^2+(\rho_{xy}^{tot}-|\boldsymbol\rho^{tot}|\nu e^2/h)^2},
}
\end{equation}
where $\boldsymbol\rho^{tot}$ is the total resistivity tensor. $|\boldsymbol\rho^{tot}|$ is the matrix determinant, $|\boldsymbol{\rho^{tot}}|=\mathrm{det}(\boldsymbol\rho^{tot})$. By Taylor expanding Eq.(\ref{equ:RxySdH}) for small $\sigma_{xx}^{Q}$ we can get a sufficient condition for the small longitudinal sheet conductivity limit for the high mobility layer where Eq.(\ref{equ:n||}) is valid:
\begin{equation}
\label{equ:condition}
{
\rho_{xx}^{tot}<\rho_{xy}^{tot} \left(1-\frac{\rho_{xy}^{tot}}{h/\nu e^2} \right)
}.
\end{equation}

We can now define the SdH regime for characterizing parallel conduction. In the data of Fig. 3, both $\rho_{xx}^{tot}$ and $\rho_{xy}^{tot}$ are shown for all temperatures below 17 K where oscillatory behavior is observed.  For $\nu =1$, Eq.~(\ref{equ:condition}) is satisfied only at 1.6 K, confirming that this data is in the QH regime, and all higher temperatures from 3.8 K to 17 K are in the SdH regime. 
\begin{figure}
	\includegraphics[width=\columnwidth]{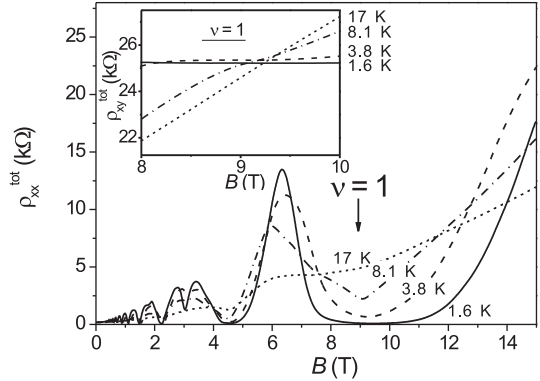}
    \caption{Measured $\rho_{xx}^{tot}$ between 1.6 K and 17 K. $\rho_{xx}^{tot}$ minima increase as temperature increases. Between 1.6 K and 20 K is the Shubnikov de Hass (SdH) regime as discussed in text.}
	\label{fig:Rxx}
\end{figure}

\begin{figure}
	\includegraphics[width=\columnwidth]{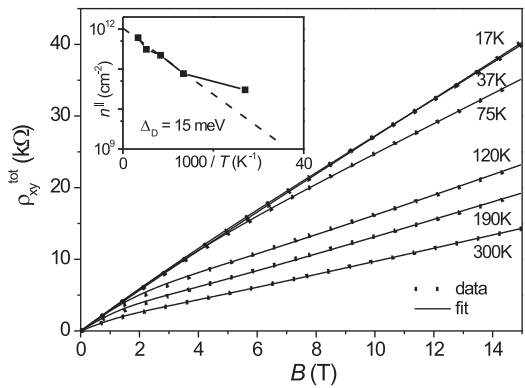}
    \caption{ Magnetic field dependence of Hall resistance at temperatures above 20 K in the classical regime. Non-linear curve fitting is adopted to extract parameters for two parallel conducting layers. Experimental data is displayed in dots and fitted results in solid line. The inset shows the possible activation energy ($\Delta_D$) for the parallel conducting channel at high temperature above 75 K. The error bars for $\Delta_D$ is $\pm 2 \, \mathrm{meV}$.
     }
	\label{fig:Rxy}
\end{figure}
Above 20 K there is no oscillatory behavior in the $\rho_{xy}^{tot}$ traces of Fig.~\ref{fig:Rxy}, however the nonlinear behavior reveals parallel conduction.
A parallel conduction Drude model \cite{Kane} can be applied to extract the conduction parameters for both the high mobility layer and the parallel channel. Separate sheet resistivity tensors for two layers are defined as
 \begin{equation}
 \label{equ:rhoi}
{\boldsymbol\rho^i = \left ( \begin{array}{rl} \rho_{xx}^i & \rho_{xy}^i \\ -\rho_{xy}^i & \rho_{xx}^i \end{array} \right ) \quad (i = Q,||),}
 \end{equation}
 where
\begin{equation}
\label{equ:Rxx,Rxy}
{\rho_{xx}^i = \frac{1}{n^ie\mu^i} =\frac{1}{\sigma_0^i}, \quad \rho_{xy}^i=\frac{B}{n^ie}= \frac{B\mu^i}{\sigma_0^i},}
\end{equation}
and $n^i$,$\mu^i$ and $\sigma_0^i$ are the sheet carrier density, mobility and sheet conductivity at zero magnetic field for layer $i$, respectively. The corresponding conductivity tensor obtained by inverting Eq.(\ref{equ:rhoi}) is
\begin{equation}
\label{equ:sigmai}
{
\boldsymbol\sigma^i=(\boldsymbol\rho^i)^{-1} =
\frac{\sigma_0^i}{1+(\mu^iB)^2}\left( \begin{array}{cc} 1 & -\mu^iB \\ \mu^iB & 1 \end{array} \right).
}
\end{equation}
The total conductivity tensor of the system is
\begin{equation}
\label{equ:sigmatot}
{
\boldsymbol\sigma^{tot}=\boldsymbol\sigma^Q+\boldsymbol\sigma^{||}.
}
\end{equation}
Inverting $\boldsymbol\sigma^{tot}$ gives the total resistivity tensor $\boldsymbol\rho^{tot}=(\boldsymbol\sigma^{tot})^{-1}$, whose components $\rho_{xx}^{tot}$ and $\rho_{xy}^{tot}$ are directly measured in a four-terminal resistance or Hall measurement.
$\rho_{xx}^{tot}$, $\rho_{xy}^{tot}$ are functions of $B$ with four parameters $\sigma_{0}^{Q}$, $\sigma_{0}^{||}$, $\mu^{Q}$ and $\mu^{||}$. In particular when $B\to0 \, \mathrm{T}$,
\begin{equation}
\label{equ:Rxx0}
{
\sigma_{0}^{Q}+\sigma_0^{||}=\frac{1}{\rho_{xx}^{tot}|_{B=0}}
}
\end{equation}
\begin{equation}
\label{equ:RH0}
{
\sigma_0^Q\mu^Q+\sigma_0^{||}\mu^{||}=\frac{d\rho_{xy}^{tot}/dB}{(\rho_{xx}^{tot})^2}|_{B=0}
}
\end{equation}
These are the well-established expressions for mixed conduction in low magnetic fields \cite{Davies}. With constraints given in Eq.~(\ref{equ:Rxx0}), (\ref{equ:RH0}), a standard curve fitting program such as Auto2Fit \cite{auto2fit} is used. Then carrier density is calculated $n^i=\sigma_{0}^i/e\mu^i $.
\begin{figure}
	\includegraphics[width=\columnwidth]{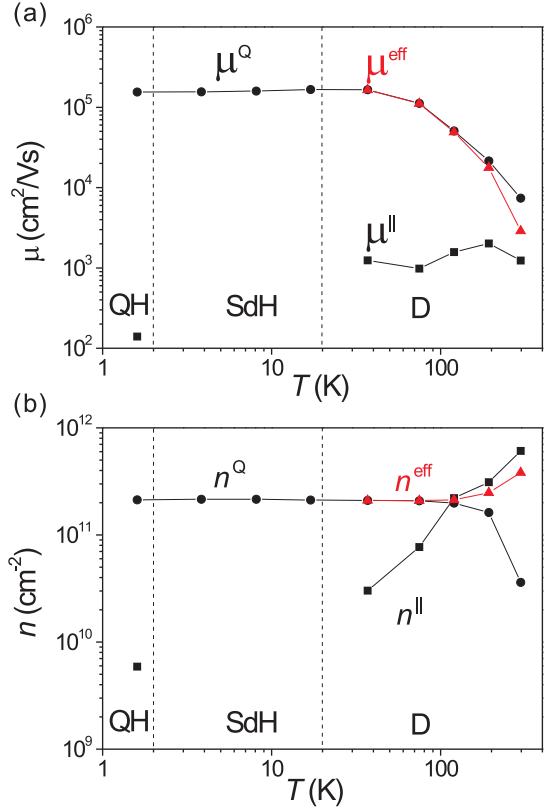}
    \caption{(a) Temperature dependence of Hall mobility.
    (b) Temperature dependence of carrier density.
    A quantum Hall (QH) parallel conduction model is used at 1.6 K when 2DEG satisfies the QH limit in strong magnetic field (QH regime) and a Drude (D) two-layer parallel conduction model is used above 20 K (D regime). The temperature in between is the SdH regime where the parallel conducting layer is impossible to characterize. The dash line marks the temperature boundary of three regimes. Parameters with superscript \emph{eff} are the values from the traditional single conducting layer model.
    }
	\label{fig:VT}
\end{figure}

Figure~\ref{fig:VT}(a) summarizes the temperature dependence of Hall mobility and Fig.~\ref{fig:VT}(b) carrier density in the entire temperature range from 1.6 K to 300 K.
 Because below 20 K the parallel channel almost freezes out, $n^{Q}$ and $\mu^{Q}$ are estimated from the sheet resistivity and Hall measurement results in the zero $B$ limit using a single conducting layer approximation. 
 The values of carrier density and Hall mobility from the two models are consistent and reveal reasonable trends for each parameter as temperature changes. $\mu^{Q}$ is almost constant of order of $10^5 \, \mathrm{cm}^2/\mathrm{Vs}$ below 40 K, and decreases rapidly with increasing temperature above 40 K to circa $10^4 \, \mathrm{cm}^2/\mathrm{Vs}$ at 300 K. $\mu^{||}$ is temperature independent from 40 K to 300 K. $n^{Q}$ is almost constant around $2 \times 10^{11} \mathrm{cm}^{-2}$, only decreasing a little from 200 K to 300 K. And $n^{||}$ increases very drastically as temperature increases.
 
The triangles in Fig.~\ref{fig:VT} represent the effective mobility and carrier density estimated from a naive single-layer conductivity model. This model is typically applied in the literature to estimate sample mobility, and it attributes all conduction to a single layer.  With this simple model, the room temperature 2D mobility underestimates the actual 2D mobility by a factor of 2.6 and carrier density underestimated by a factor of 1.6. As temperature decreases below 140 K, however, there is no appreciable difference between the single layer approximation and the estimates of the high mobility layer from the parallel conduction model.

The data reported here testify to the high growth quality of the samples.  The temperature dependence of 2D mobility has been published by Matsuoka. {\itshape et. al.} \cite{Matsuoka} for InGaAs/InAlAs heterostructures with a saturation mobility about $6 \times 10^{4} \, \mathrm{cm}^2/\mathrm{Vs}$ and a knee at 77 K.  By comparison, the mobility here saturates at $1.6 \times 10^{5} \, \mathrm{cm}^2/\mathrm{Vs}$ and shows a knee 40 K. The knee in $\mu^{Q}$ is known to be be caused by polar LO phonon scattering, and the saturation at low temperatures is interpreted in terms of alloy disorder scattering \cite{Chatto}. The higher saturation mobility and lower knee temperature indicate that our sample has less alloy disorder scattering and better quality. 

The analysis shown here also allows other energy scales in the band diagram to be deduced, such as the dopant binding energy, and the heterojunction conduction band offset.  The activation energy of the parallel conduction states can be estimated by fitting an Arrehnius law $n^{||}=n_0\exp (-\Delta_D/k_BT)$ above 75 K as shown in the inset of Fig.~\ref{fig:Rxy}, where $\Delta_D$ is the thermal activation energy, $k_B$ the Boltzmann constant, $n_0$ represents the number density of impurities. Average $\Delta_D$ is about 15 meV and $n_0$ is around $10^{12} \, \mathrm{cm}^{-2}$.

The conduction band offset ($\Delta E_C$) at the base temperature 1.6 K can be estimated with the parameters extracted above. Since the assumed background dopant density of InP bulk ($\sim 10^{15} \, \mathrm{cm}^{-3}$) is too low to offer enough electrons for 2DEG, it is reasonable to assume that InP layer is entirely depleted and the main source of donor is interfacial states between InP epilayer and semi-insulating InP substrate. This assumption leads to a conduction band structure as shown in panel (b) of Fig.~\ref{fig:structure}. It is clear that
\begin{equation}
\label{equ:bandoffset}
{
\Delta E_C=ed{\Large \mbox {$\varepsilon$}} _B + \Delta _D + E_F+\Delta_0
},
\end{equation}
where $d=200 \, \mathrm{nm}$ is the thickness of InP bulk, ${\Large \mbox {$\varepsilon$}} _B$ the electrical field strength in InP, $\Delta _D$ the binding energy of interfacial states, and $\Delta_0$ the activation energy from the bottom of the quantum well to the first subband. According to Maxwell's equations,
\begin{equation}
\label{equ:EB}
{
\epsilon _r^B \epsilon _0 {\Large \mbox {$\varepsilon$}} _B = \epsilon _r^W \epsilon _0 {\Large \mbox {$\varepsilon$}} _W = n^{Q}e
},
\end{equation}
where $\epsilon _r^B= 12.4$ and $\epsilon _r^W = 14$ are the relative static permittivity for InP and In$_{0.532}$Ga$_{0.467}$As respectively \cite{Guldner}, $\epsilon _0$ the electrical constant and ${\Large \mbox {$\varepsilon$}} _W$ the electrical field in the quantum well. Then with $n^{Q} = 2.1 \times 10^{11} \, \mathrm{cm}^{-2}$ the conduction band decrease in InP bulk can be calculated:
\begin{equation}
\label{equ:edEb}
{
ed{\Large \mbox {$\varepsilon$}} _B = \frac{n^{Q}e^2d}{\epsilon _r^B \epsilon _0}= 612 \, \mathrm{meV}.
}
\end{equation}
The interface dopant density is $n_0=10^{12} \, \mathrm {cm}^{-2}$  and $\Delta _D$ is 15 meV as mentioned above.
$E_F$ can also be calculated from $n^{Q}$ as
\begin{equation}
\label{equ:Ef}
{
E_F = \frac{ \pi \hbar ^2}{m_e^*}n^{Q}\approx  12.3 \, \mathrm{meV} ,
}
\end{equation}
where the effective electron mass in 2DEG is approximated by that in the bulk of In$_{0.532}$Ga$_{0.467}$As, $m_e*\approx 0.041m_0$. Finally we can estimate $\Delta _0$ with the model developed by Stern \cite{Spern}
\begin{equation}
\label{equ:delta0}
{
\Delta _0 = \frac{5}{8}\left(33e^2 \hbar n^{Q} / 8 \epsilon _0 \epsilon_r^W {m_e^*}^{1/2}\right)^{2/3}.
}
\end{equation}
With $\Delta_0 \approx 38 \, \mathrm{meV}$, the conduction band offset can be estimated as $\Delta E_C \approx 677 \, \mathrm{meV} $, about 140 meV larger the result obtained by Guldner \emph{et. al.} in similar heterostructure \cite{Guldner}. In addition to possible differences in alloy content, the difference in $\Delta E_C$ may also be caused by either different interface defects with differing binding energy, and/or different background bulk dopant density which would add minor curvature to the conduction band not included in the above approximation.

In summary, we have presented the magnetotransport results for the high quality InP/InGaAs heterostructure and characterized them with parallel conduction models in different temperature ranges. The mobility and carrier density for 2DEG in the entire temperature range between 1.6 K and 300 K are extracted. So are those parameters for the parallel channel in the classical and quantum Hall regime. The main source of donors is believed to be the interfacial states between InP and substrate. The methods used in this paper can be applied to the general characterization of heterostructures with parallel conduction of layers with different mobilities and densities.

\begin{acknowledgments}
The authors thank Stanley Tsao and M. Razeghi for sample growth and preparation.

This work was supported by the MRSEC program of the National Science Foundation (DMR-0520513) at the Materials Research Center of Northwestern University and AFOSR Grant FA9550-09-1-0237.
\end{acknowledgments}

\end{document}